%% file: main.tex
\newif\ificlr
 \iclrfalse     

\documentclass{article}

\ificlr
    \usepackage{iclr2027_conference,times}
\else
    \usepackage[margin=1in]{geometry}
    \usepackage{times}
    \usepackage{authblk}
    \usepackage{natbib}
\fi

\usepackage{amsmath,amssymb,booktabs,multirow,graphicx,microtype,url,xcolor}
\providecommand{\nolinkurl}{\url}
\usepackage[T1]{fontenc}
\usepackage[utf8]{inputenc}

\newcommand{\RR}{\mathbb{R}}
\newcommand{\Loss}{\mathcal{L}}
\newcommand{\Dval}{\mathcal{D}_{\mathrm{val}}}

\newcommand{\pkg}{\textsc{LookAgain-ML}}

\input{generated/scalars}
\input{generated/nn_scalars}
\input{generated/flowers_scalars}

\title{Representation Risk in Pretrained Image Encoders}

\ificlr

    \author{Anonymous Authors\\
    Paper under double-blind review}

\else

    \author[1]{Ardyn Nordstrom}
    \author[2]{Morgan Nordstrom}
    \author[1]{Vamuyan Sesay}
    \author[3]{Matthew D. Webb}

    \affil[1]{School of Public Policy and Administration, Carleton University}
    \affil[2]{Independent Researcher}
    \affil[3]{Department of Economics, Carleton University}

    \date{\today}

\fi
\begin{document}
\raggedbottom
\maketitle

\begin{abstract}
Applied researchers increasingly convert images into features with pretrained encoders, then use those features in a downstream prediction model. The encoder is often treated as an implementation detail. We show that it can instead be a consequential source of model uncertainty. We call this uncertainty \emph{representation risk}: plausible pretrained encoders map the same images into different feature spaces and can yield sharply different out-of-sample conclusions from predictive performance. We compare ten modern and legacy frozen encoders across applications involving house prices, racehorse performance, breast-cancer histology, chest radiographs, continuous facial age, and rice disease. With common dimension control, heads, and group-safe splits, validation selects SigLIP~2 for houses, raising test $R^2$ from \HouseResNet{} for ResNet50 to \HouseSelected{}, and DINOv2 for horses, raising $R^2$ from \HorseResNet{} to \HorseSelected{}. No encoder is best in every task. Candidate procedures are constructed using training data and compared on a separate validation partition. The selected procedure reaches \HousePreferred{} for houses and \PneumoniaPreferred{} accuracy for pneumonia. Fixed-split gains are small for horses and rice, while repeated partitions reveal instability in horse feature union. Continuous age selects SigLIP~2 at \AgeSelected{} years MAE. The principal representation gaps persist with neural heads, similarly sized DINOv2 and ViT models, and limited adaptation. These results support a simple workflow: benchmark plausible representations, select on locked validation data, combine only when separate validation evidence justifies the additional cost, and report paired and split-level uncertainty. We implement this workflow in \pkg{}, the software package used to conduct the analyses in this paper.
\end{abstract}

\section{Introduction}
\label{sec:introduction}

The same photograph of a Toronto house explains substantially different amounts of out-of-sample price variation depending only on which pretrained image encoder converts the photograph into features. With a frozen ResNet50 representation and a ridge head, test $R^2$ is \HouseResNet{}. Validation selects SigLIP~2 from the ten-encoder menu fixed in the implemented workflow, raising $R^2$ to \HouseSelected{} while keeping the sample, grouping and split, preprocessing protocol, head family, and tuning rule common. A validation-selected linear stack reaches \HousePreferred{}. The first change is more important than the second: choosing the representation accounts for most of the improvement, while combination supplies an additional gain only after strong representations have been identified.

An image encoder maps a raw image into a numerical feature vector, or \emph{representation}, that is subsequently used for prediction. Pretrained encoders make image analysis feasible in applications with limited labeled data, and the resulting vectors can be joined to outcomes and covariates like ordinary tabular features. Yet an applied workflow must still choose among architectures, training objectives, pretraining corpora, and output conventions. That choice is often inherited from old code or made without a direct comparison. We use \emph{representation risk} to describe the resulting uncertainty: different plausible pretrained representations of the same images can lead to different downstream performance and, therefore, different empirical conclusions. Representation learning is standard terminology; our term names an applied model-selection problem rather than a new learning paradigm.

We examine how large and stable representation risk is when modern and legacy encoders are compared under a common downstream protocol, and when combining representations can mitigate it. Diversity alone cannot resolve the combination problem. Two weak encoders may disagree because both are noisy. Combination is useful when encoders are individually informative and their remaining errors are complementary. Houses illustrate that regime: the selected stack has a paired $R^2$ gain of \HouseGain{} over SigLIP~2 (95\% CI $[\HouseGainLower{},\HouseGainUpper{}]$), which was previously examined in a working paper by \citet{nordstrom2024using}. Standardized racehorse conformation photographs from the Ocala Breeders’ Sales (OBS) illustrate the contrast. DINOv2 raises workout-speed $R^2$ from \HorseResNet{} to \HorseSelected{}, while the selected feature union reaches \HorsePreferred{}; its gain over DINOv2 is \HorseGain{} (95\% CI $[\HorseGainLower{},\HorseGainUpper{}]$). Across twelve group-safe splits, DINOv2 remains the selected horse encoder, but feature union suffers one extreme test-performance failure.

We make three contributions. First, we quantify representation risk across six applied image tasks using modern encoders---DINOv2, SigLIP~2, ConvNeXt, and ViT---and legacy convolutional and semantic representations. Encoder choice changes performance materially, and the validation-selected encoder varies across tasks. Second, we compare validation-based selection, equal prediction averaging, linear stacking, feature union, and sparse subset selection. Candidate procedures are constructed within training data and assessed on a separate validation partition; combination helps in some tasks, but no rule is a reliable default. Third, we introduce \pkg{}, the software package used for all analyses in this paper. It provides a common workflow for caching pretrained representations, applying common prediction models and group-safe splits, selecting and combining encoders using validation
data, measuring uncertainty and computational cost, and recording the models required for
inference. Neural-head, transferability, adaptation, dimension, capacity, and near-duplicate checks probe the main representation gaps.

\begin{figure}
    \centering
     \caption{Representation choice is first-order; model combination is second-order}
    \includegraphics[width=0.99\linewidth]{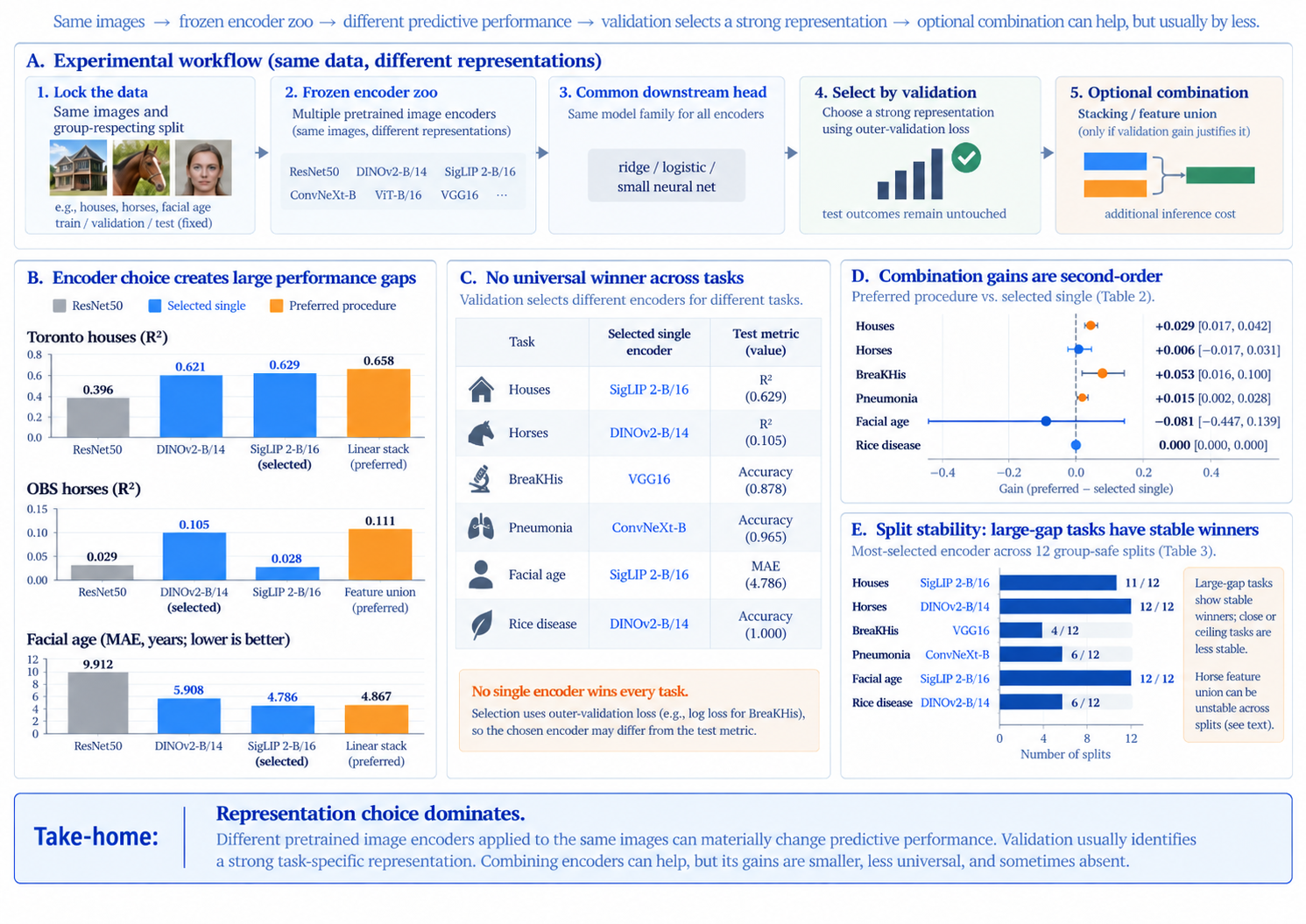}
   
    \label{fig:overview}
\end{figure}

Figure~\ref{fig:overview} summarizes the paper's main argument and empirical findings.
Holding the images and prediction task fixed, encoder choice can generate large differences
in out-of-sample performance, and the strongest encoder varies across tasks.
Validation generally identifies a strong task-specific representation, while combining
representations produces smaller and less consistent gains.

\section{Related Work}
\label{sec:related}

\paragraph{Pretrained representations and model selection.}
Supervised ImageNet models such as VGG, ResNet, Inception, and MobileNetV2 established the common practice of transferring intermediate image features \citep{russakovsky2015imagenet,simonyan2015very,he2016deep,szegedy2016rethinking,sandler2018mobilenetv2}. Vision Transformers and ConvNeXt updated the architecture, while contrastive and self-supervised objectives produced CLIP, DINO, DINOv2, and SigLIP~2 \citep{dosovitskiy2021image,liu2022convnet,radford2021learning,caron2021emerging,oquab2023dinov2,tschannen2025siglip2}. A related literature estimates transferability before full task training. Task2Vec embeds tasks, LEEP estimates transferability from a pretrained classifier's outputs, and LogME evaluates the evidence of features for a target \citep{achille2019task2vec,nguyen2020leep,you2021logme}. \citet{you2022ranking} extend this line to ranking the models in a model hub and
to tuning with several top-ranked models. ZooD exploits a model zoo for out-of-distribution generalization \citep{dong2022zood}. We study a complementary applied decision: after plausible encoders are available, how much do their locked-split results differ, and is selection or combination warranted?

\paragraph{Ensembles and feature fusion.}
Bagging, stacking, and ensemble learning are established methods \citep{breiman1996bagging,wolpert1992stacked,dietterich2000ensemble}. Likewise, early, intermediate, and late fusion are standard ways to combine heterogeneous features or predictions \citep{atrey2010multimodal}. We do not propose averaging or stacking as new estimators. Our contribution is empirical and procedural: we separate the consequences of choosing a pretrained representation from the conditional value of combining several representations, use validation-locked decisions, and connect gains to strength, error complementarity, and inference cost.

\paragraph{Images in applied prediction.}
Images have been used to measure poverty, neighborhood conditions, housing quality, and latent attributes that are difficult to code by hand \citep{jean2016combining,yeh2020using,law2019take,glaeser2018computer}. Recent real-estate work combines images with structured, textual, or geospatial inputs \citep{yazdani2023real,hasan2024multi,yousif2023real}. CityLens benchmarks modern vision-language systems for region-level urban socioeconomic sensing, including a house-price task \citep{liu2026citylens}; a recent CLIP-based quality score studies a much larger property-image corpus \citep{slonimczyk2026clip}. Our house application differs in unit and purpose: it predicts property-level sale prices while holding the sample, grouping and split, head family, preprocessing protocol, and tuning rule common across encoders; fitted heads and selected penalties can differ. More broadly, we compare encoder choice across economic, medical, biological, and demographic applications rather than designing a domain-specific multimodal system.

\paragraph{Transfer variation, underspecification, and analytic flexibility.}
The observation that encoder rankings depend on the downstream task predates current foundation models. \citet{kornblith2019better} find that training choices that raise ImageNet accuracy do not always improve fixed-feature transfer, and the Visual Task Adaptation Benchmark shows that representations performing well on natural-image tasks need not perform well on specialized or structured tasks \citep{zhai2019vtab}. These studies benchmark classification transfer on standard splits. We differ from this work by studying modern encoders that already lead such benchmarks, on applied regression and classification tasks with group-level leakage control, and ask what locked validation selection and combination recover. Our framing is also related to work on underspecification and predictive multiplicity, which emphasizes that multiple models can achieve similar predictive performance while differing in important ways. Pipelines with equivalent held-out performance can differ in deployment behavior \citep{damour2022underspecification}, and models with near-equal accuracy can assign conflicting individual predictions \citep{marx2020predictive}, a phenomenon \citet{breiman2001statistical} called the Rashomon effect. Representation risk differs in that the competing representations are not near-equivalent: the variation is in the level of
held-out performance itself. It is closer to the analytic flexibility studied by multiverse and specification-curve analysis, in which several defensible analysis choices are reported jointly rather than one being fixed silently
\citep{steegen2016multiverse, simonsohn2020specification}.

\section{Framework}
\label{sec:framework}

Let $x_i$ be an image and $y_i$ its prediction target. A frozen encoder $e_m$ maps the image to $z_{im}=e_m(x_i)\in\RR^{d_m}$. A downstream head $h_m$ produces $\hat y_{im}=h_m(z_{im})$. Within a dataset, we hold the observations, train/validation/test split, preprocessing discipline, and head family fixed across $m$. Differences in held-out loss therefore measure the practical consequence of representation choice, including how readily a common head can extract task-relevant information.

We use \emph{representation risk} operationally, not as a new population risk functional. It is the sensitivity of downstream performance to the choice among defensible pretrained encoders. We report the spread of single-encoder performance, paired differences between important encoders, and whether the validation-selected encoder changes across tasks. This definition deliberately includes linear accessibility. Section~\ref{sec:nn} tests whether the conclusions persist when every representation receives a nonlinear head.

\paragraph{Selection and combination.}
For a validation loss $\Loss$, single-model selection chooses
\begin{equation}
  \widehat m=\arg\min_m\sum_{i\in\Dval}\Loss\!\left(y_i,\widehat y_{im}\right).
\end{equation}
Equal averaging uses $M^{-1}\sum_m\widehat y_{im}$ for regression and averages class probabilities for classification. Linear stacking estimates an intercept and weights; multinomial stacking operates on class-probability vectors. We also concatenate standardized encoder blocks (feature union) and use greedy forward selection to construct sparse averages. These candidates share one evaluation boundary. Within outer training, group-safe cross-fitting tunes each base head and produces out-of-fold predictions. Stack weights and the greedy subset use only those predictions and outer-training outcomes. Feature-union scaling, principal components, and head tuning use a group-safe inner partition of outer training. The fitted candidates are frozen before their losses are compared on outer validation. Equal averaging has no fitted parameters.

After the preferred procedure is selected, all decisions are locked. Base heads are refitted on training plus validation with their chosen hyperparameters. A selected stack is refitted from new group-safe out-of-fold predictions on training plus validation; a selected subset keeps its sequence fixed; feature union is refitted using its locked inner choice. Test outcomes are excluded from fitting and model selection. Classification selection uses validation log loss even when tables display accuracy; age selection uses MAE in years, while houses and horses use MSE. Test outcomes do not fit, select, clip, or stabilize any candidate.

The central distinction is between \emph{strength} and \emph{complementarity}. An encoder can disagree with a strong model because it contains additional signal or because it is inaccurate. We therefore report standalone performance together with pairwise correlations of predictions and errors. For classification, our primary error-diversity measure is the correlation of per-observation log loss; misclassification-indicator correlations are a robustness measure. These are diagnostics, not a universal selection formula.

\paragraph{Uncertainty.}
We form paired 95\% intervals by resampling the locked test groups 2,000 times and recomputing both performance metrics and their difference from fixed fitted predictions. Models are not retrained inside a bootstrap draw. Resampling respects the saved leakage groups for each task. The intervals condition on the fitted models and locked split; twelve group-safe splits fixed in the implemented workflow provide a separate stability summary. Neural-head metrics are based on predictions averaged across five prespecified optimization seeds.

\section{Experimental Design}
\label{sec:design}

Primary analyses in the paper are conducted using \pkg{}. Table~\ref{tab:datasets} summarizes the six main tasks. Houses use the saved leakage-group partition, which coincides with exact image-file-hash grouping in the analyzed sample. Horses use 9,461 reconstructed identity groups across 9,729 rows, joining exact normalized names, complete pedigree--sex--foaling identifiers, and photograph hashes transitively. BreaKHis and chest X-ray use patient-aware splits \citep{spanhol2016dataset,kermany2018identifying}; exact duplicate images are grouped for facial age and rice disease \citep{sethy2020deep}.  For datasets without a defensible pre-existing split, we use one locked 70/15/15 split and twelve alternative group-safe splits fixed in advance.

\begin{table}[t]
\centering
\small
\setlength{\tabcolsep}{4.6pt}
\caption{Main datasets and locked test designs. $N$ is the common-complete sample used for encoder comparisons.}
\label{tab:datasets}
\begin{tabular}{lrrll}
\toprule
Dataset & $N$ & Test $N$ & Task & Primary metric\\
\midrule
Toronto houses & 7,273 & 1,090 & log sale price & $R^2$\\
OBS horses & 9,729 & 1,459 & normalized workout speed & $R^2$\\
BreaKHis & 7,909 & 1,129 & binary histology & accuracy\\
Pneumonia & 5,856 & 878 & binary chest X-ray & accuracy\\
Facial age & 9,764 & 1,465 & chronological age & MAE (years)\\
Rice disease & 5,932 & 759 & four leaf classes & accuracy\\
\bottomrule
\end{tabular}
\end{table}

We compare ResNet50, VGG16, InceptionV3, MobileNetV2, and semantic features derived from COCO and ADE20K with DINOv2-B/14, SigLIP~2-B/16, ConvNeXt-B, and ViT-B/16. Encoders are frozen, and cached representations are reused in downstream analyses. The primary design applies train-only PCA with a 128-dimensional cap before a standardized ridge or regularized logistic head. Representations with fewer than 128 coordinates retain their native dimension. We report $R^2$, MSE, MAE, correlation, and calibration slope for regression; facial age additionally reports the percentage within five years. Classification summaries include accuracy, balanced accuracy, macro F1, log loss, AUC for binary tasks, and expected calibration error where appropriate. Additional metrics are available in the accompanying results files.

Automated checks verify image--label and cache alignment, finite arrays and probabilities, zero group/hash overlap across fixed and repeated splits, train-only scaling and PCA, group-safe out-of-fold meta-model fitting, and untouched test outcomes. Embedding caches without recoverable row identifiers were excluded. The analysis uses semantic class-share features from a DeepLab model trained on COCO. Near-null directions use a fixed float64, training-only floor; no test-informed clipping or winsorization is used.

\section{Results}
\label{sec:results}

\subsection{Representation choice changes prediction}

\begin{figure}[t]
\centering
\includegraphics[width=\textwidth]{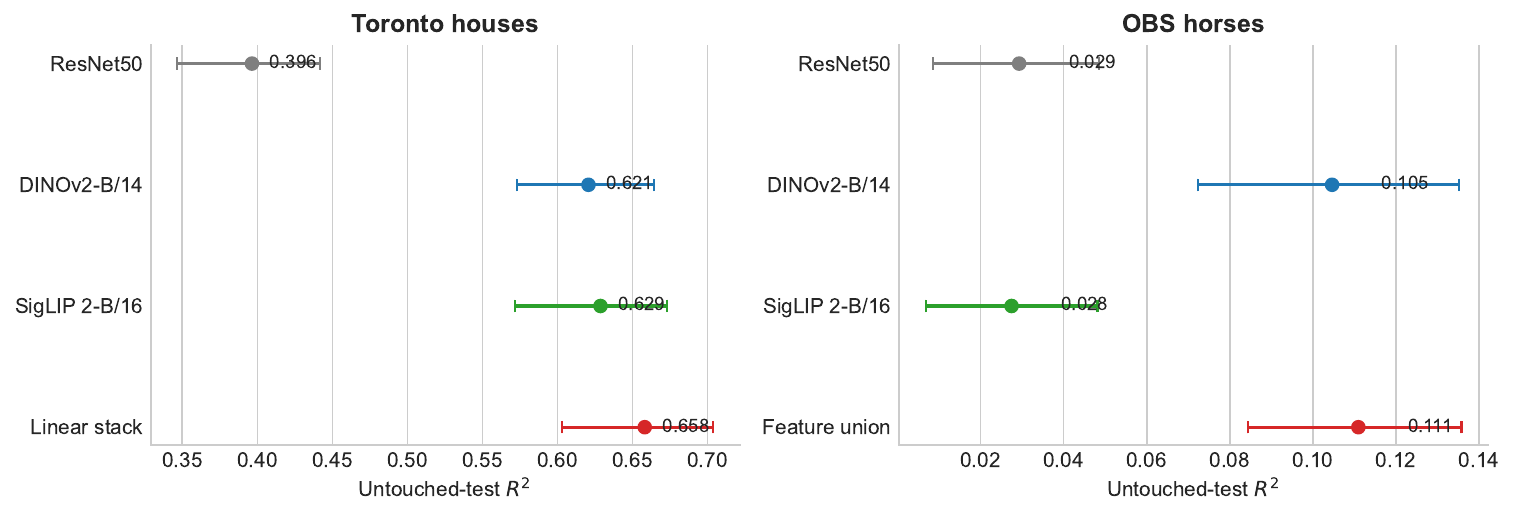}
\caption{\textbf{Representation choice and combination in the two motivating applications.} Dots report locked-test $R^2$; horizontal bars are 95\% grouped-bootstrap intervals from fixed fitted predictions. The same observations, dimension control, and ridge-head protocol are used within each panel. The final row is the procedure selected on outer-validation loss: a linear stack for houses and feature union for horses.}
\label{fig:motivation}
\end{figure}

Figure~\ref{fig:motivation} shows the two clearest examples. For houses, validation-selected SigLIP~2 raises $R^2$ from \HouseResNet{} to \HouseSelected{}. For horses, DINOv2 raises explained variation from \HorseResNet{} to \HorseSelected{}. The point is not that either encoder always wins. Validation selects different encoders across tasks, and the untouched-test leader can differ from the validation choice. A familiar default can therefore be far from the strongest available representation, while small validation differences can also make selection noisy.

\begin{figure}[t]
\centering
\includegraphics[width=\textwidth]{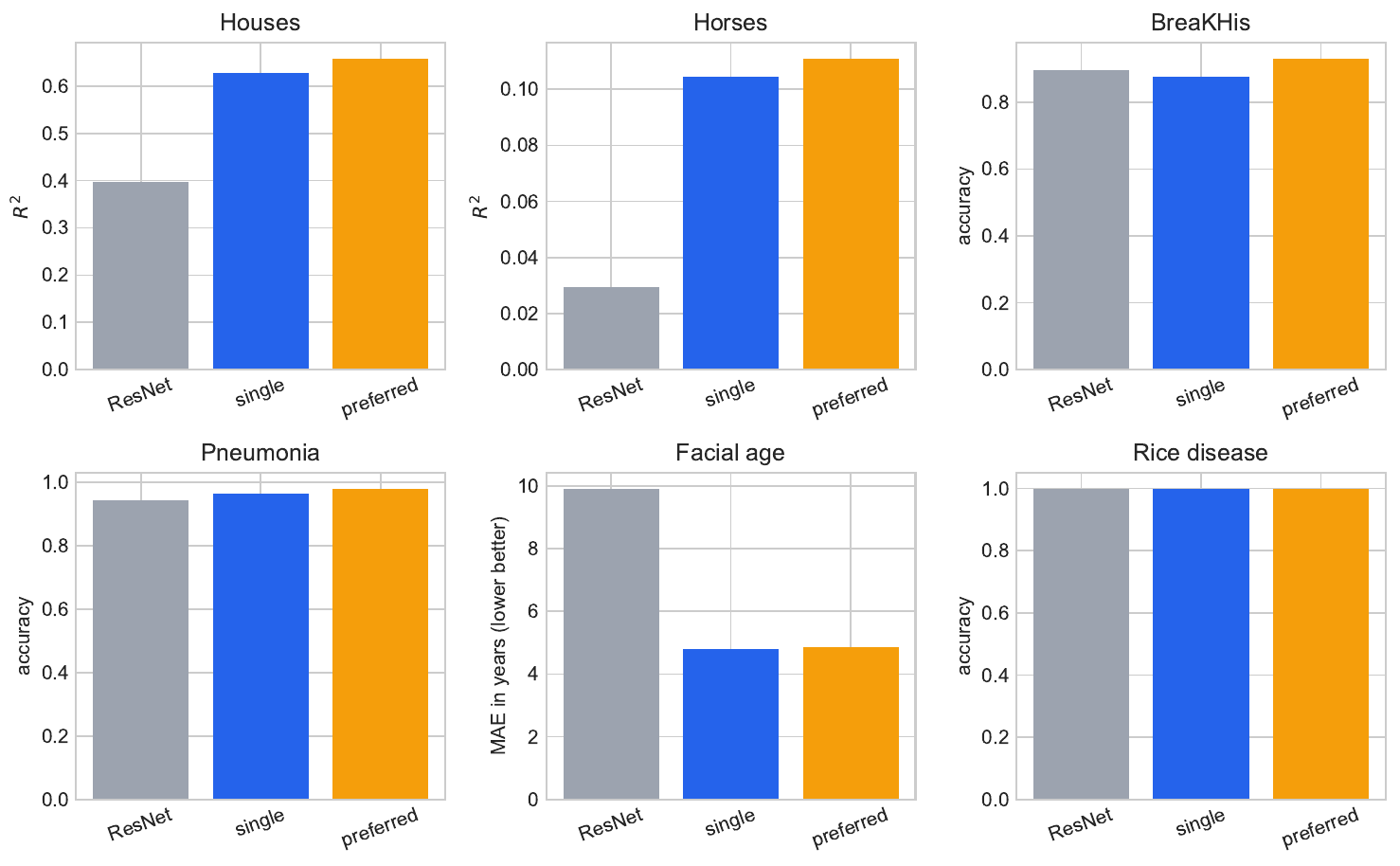}
\caption{\textbf{Encoder choice and preferred procedure across the six main tasks.} Houses and horses report $R^2$; BreaKHis, pneumonia, and rice report accuracy. Facial age is shown on its own MAE-in-years scale, where lower is better. Selection uses validation MSE, MAE, or log loss as appropriate, not the displayed test metric.}
\label{fig:crossdomain}
\end{figure}

The cross-domain comparison in Table~\ref{tab:headline} and Figure~\ref{fig:crossdomain} rejects a universal encoder recommendation. Validation selects SigLIP~2 for houses and continuous age, DINOv2 for horses and rice, VGG16 for BreaKHis, and ConvNeXt for pneumonia. The untouched-test leader sometimes differs, but test outcomes never alter the locked choice. BreaKHis makes the selection-metric distinction particularly visible: VGG16 minimizes validation log loss even though its printed test accuracy is below several alternatives. Rice is at a ceiling, so neither representation choice nor combination has room to improve test accuracy.

\begin{table}[t]
\centering
\small
\setlength{\tabcolsep}{3.6pt}
\caption{Locked-test headline results. The selected single minimizes outer-validation loss over all ten encoders. The preferred procedure minimizes comparable outer-validation loss over that single and the combination menu fixed in the implemented workflow. Gain is oriented so positive values favor the preferred procedure.}
\label{tab:headline}
\resizebox{\textwidth}{!}{\input{tables/generated_headline}}
\end{table}

\subsection{Combination helps conditionally}
\label{sec:combination}

The house and BreaKHis results show that combination can improve on a selected single representation. For houses, a training-OOF linear stack raises $R^2$ from \HouseSelected{} for SigLIP~2 to \HousePreferred{}. Its paired gain is \HouseGain{} (95\% CI $[\HouseGainLower{},\HouseGainUpper{}]$). For BreaKHis, equal probability averaging raises accuracy from \BreakHisSelected{} for validation-selected VGG16 to \BreakHisPreferred{}, with a paired interval excluding zero. These are gains over the locked selection benchmark, not comparisons chosen from test performance.

Other tasks show why this is not a general averaging result. The selected horse feature union gains only \HorseGain{} over DINOv2, with an interval spanning zero, and requires several encoder passes. On the fixed age split, the selected stack increases MAE by 0.081 years, and its interval spans zero. Rice has no remaining classification errors to diversify. Pneumonia provides a second positive case: selected feature union raises accuracy from \PneumoniaSelected{} to \PneumoniaPreferred{}. Complementarity is useful only among representations that already contain signal, and outer-validation preference does not guarantee a test improvement.

Equal averaging and feature union are similarly conditional. Equal averaging helps BreaKHis but dilutes strong models in several other tasks. Feature union is selected for pneumonia and horses, but performs poorly for houses. Validation-based selection therefore matters twice: it protects against an arbitrary single encoder, then limits combination to cases where a separately assessed validation loss supports the extra complexity.

\subsection{Split stability separates large gaps from close calls}
\label{sec:stability}

Table~\ref{tab:repeated} summarizes twelve group-safe splits fixed in the implemented workflow. Houses gain about 0.030 $R^2$ on average from the preferred procedure. SigLIP~2 is selected for continuous age in all twelve splits; mean selected-single MAE is \AgeRepeatedSelected{} and mean preferred-procedure MAE is \AgeRepeatedPreferred{} with split SD \AgeRepeatedSD{}. BreaKHis most often selects VGG16 (4/12). Pneumonia gains about 0.018 accuracy on average, while rice remains at the ceiling. Test-best regret is retained only as a diagnostic and never affects selection.

\begin{table}[t]
\centering
\scriptsize
\caption{Twelve group-safe repeated splits. Age columns are MAE; other performance columns use $R^2$ or accuracy. Gains are oriented so positive values favor the preferred procedure: preferred minus selected-single performance for $R^2$ and accuracy, and selected-single minus preferred MAE for age. ``Preferred SD'' is the sample standard deviation of preferred-procedure test performance across the twelve partitions, not a standard error. Means and differences are computed before rounding. ``Most selected'' reports all encoders tied at the maximum frequency.}
\label{tab:repeated}
\resizebox{\textwidth}{!}{\input{tables/generated_repeated}}
\end{table}

DINOv2 is selected for horses in all twelve splits and averages $R^2=\HorseRepeatedSelected{}$. Feature union wins outer validation in three splits. One selected feature-union fit produces test $R^2=-49.112$, indicating severe feature-union instability; another union split yields $-0.099$. The preferred-procedure mean is therefore \HorseRepeatedPreferred{}. The numerical safeguard addresses near-zero variation in the training data but does not constrain test predictions. We have not verified the precise cause of the failure. The full mean and failed split remain in the analysis, without test-informed clipping.

\subsection{Dimension, capacity, transferability, and data geometry}

The primary specification applies training-only PCA with a 128-dimensional cap; representations below the cap retain their native coordinates. The principal house, horse, and age gaps remain large under this dimension-controlled specification.

Across twelve splits, matched DINOv2-B/14 minus supervised ViT-B/16 averages \MatchedHouseGap{} $R^2$ for houses and \MatchedHorseGap{} for horses; DINOv2 reduces age MAE by \MatchedAgeGap{} years. DINOv2 also leads for pneumonia and rice, whereas ViT leads on BreaKHis. Because the models are similar in scale and output dimension, model size alone does not explain the gap; the comparison does not isolate a causal training-objective effect.

LogME has median Spearman association \LogMEMedianValidationRho{} with validation ranking and agrees with the validation winner in \LogMEWinnerAgreement{} tasks. Its BreaKHis association is weak, so LogME is a screening diagnostic rather than a replacement for outcome-based validation. Under the limited-adaptation protocol in Appendix~\ref{app:adaptation}, predictions averaged across seeds yield $R^2=\LightResNet{}$ for ResNet50 and $\LightDino{}$ for DINOv2, a paired gap of \LightGap{} (95\% CI $[0.142,0.274]$). The representation gap remains after limited adaptation.

For continuous age, \AgeDuplicateFlagged{} of \AgeDuplicateTotal{} test observations exceed the 0.99 nearest-training cosine threshold fixed in the implemented workflow; DINOv2 MAE changes from 5.908 to \AgeDuplicateMAE{} after excluding them. For rice, \RiceCloseShare{} of test images exceed 0.99 similarity, 1-NN accuracy is \RiceOneNN{}, and DINOv2 remains perfect after exclusion. Accuracy approaches one by 500 training images per class. Rice is therefore best read as an easy-manifold ceiling case.

\subsection{Representation risk survives nonlinear heads}
\label{sec:nn}

\begin{figure}[t]
\centering
\includegraphics[width=0.92\textwidth]{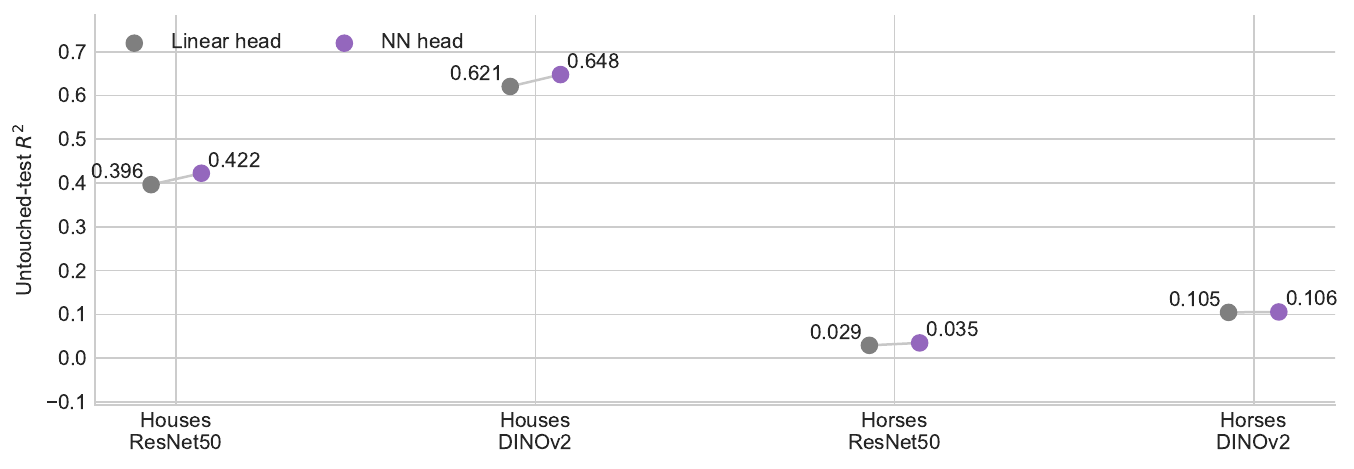}
\caption{\textbf{Flexible heads do not eliminate the ResNet50--DINOv2 gap.} Dots report locked-test $R^2$ for the common linear head and the small neural head on houses and horses. Neural metrics use predictions averaged across five seeds fixed in the implemented workflow. The full encoder comparison and the mean-rank diagnostic, which uses average ranks for ties and lower-is-better ranks, appear in Appendix~\ref{app:nnfull}.}
\label{fig:nn}
\end{figure}

One concern is that linear heads measure accessibility rather than information: perhaps older features contain the same signal in a less linearly separable form. We replace the ridge or logistic model with a small multilayer perceptron. For each dataset, we choose the network architecture using only the training data and average predictions across five prespecified seeds. Figure~\ref{fig:nn} shows the direct ResNet50--DINOv2 comparison. For houses, the NN-head DINOv2--ResNet gap remains \HouseNNGap{}. For horses, the corresponding gap is \HorseNNGap{}. Across the six tasks, the median within-task Spearman correlation between linear- and NN-head encoder rankings is \MedianNNRankCorrelation{}.

Nonlinearity helps selectively, but it does not systematically erase encoder ranking differences. These results support linear heads as the clean primary comparison and neural heads as a robustness check, not as a replacement for representation benchmarking.

\subsection{Cost and deployment}

Cost has three distinct components. Representation extraction produces features that can be cached and reused for downstream fitting. Recorded extraction measurements depend on hardware, batch size, sample, and I/O; they do not establish that extraction dominates all fitting costs. Deployment is different again: a selected single requires one encoder pass, whereas an average, stack, subset, or feature union requires every constituent encoder. The house stack improves performance while requiring multiple encoder passes; the horse feature union buys only a small, uncertain fixed-split gain. Deployment latency was not measured. The practical objective is therefore not to run every encoder by default, but to weigh validation evidence against the selected procedure's inference cost.


Whether these image predictions add information to an existing structured dataset is distinct from the choice among image representations. 
An encoder that performs well using images alone need
not provide the largest incremental gain once structured covariates are included. We therefore
hold the prediction task fixed and evaluate representation choice without using structured-model
performance to rank encoders.

\section{Discussion and Practical Recommendation}
\label{sec:discussion}

The evidence supports a simple sequence: \emph{benchmark, validate, select, optionally combine, and report uncertainty and cost}. Benchmarking addresses the first-order risk that a familiar encoder is poorly matched to the task. A separate outer-validation partition makes the procedure-selection process transparent and verifiable. Combination is a second-stage choice, justified only when strong representations improve separate validation loss enough to warrant their inference cost. Equal averaging is not a safe substitute for this decision, because weak encoders can add disagreement without useful signal.

\pkg{} implements this sequence and is used throughout the empirical analysis. It caches frozen embeddings, applies common linear and optional neural heads, maintains group-safe splits, constructs meta-models from training-only out-of-fold predictions, compares task-appropriate metrics, estimates paired intervals, and records extraction measurements and identifies the encoders required for inference. It does not eliminate model uncertainty or promise a universal winner. It makes a previously implicit encoder choice explicit and verifiable.

Several limits matter. Our dataset count is too small to establish a general law linking residual diversity to ensemble gains. Test-set bootstrap intervals condition on fixed fitted predictions, while twelve splits address only one additional source of uncertainty. Foundation-model availability and licenses change, and our encoder zoo is necessarily incomplete. Because facial-age identity labels are unavailable beyond exact image hashes, the nearest-neighbor analysis is a sensitivity check and cannot establish complete identity separation. More broadly, our contribution is a reproducible framework for comparing and selecting representations rather than a new estimator.


\section{Conclusion}

Pretrained image encoders are not interchangeable preprocessing tools. Across six applied tasks, changing the encoder while holding the data and downstream protocol common can change out-of-sample performance enough to alter the conclusions from predictive performance. No encoder wins universally. Validation-based combination provides an additional gain in some tasks, but it adds little in highly correlated or ceiling-effect settings and can fail under feature shift. Flexible neural heads do not erase the principal representation gaps. Applied image analysis should therefore treat representation choice as a reported modeling decision: compare plausible encoders, fit combinations without reusing their assessment outcomes, combine selectively, and attach uncertainty and cost to the result.

\section*{Reproducibility Statement}

All reported results were generated using the accompanying code and checked against the saved analysis outputs.
The primary analyses use \pkg{} commit \PackageCommit{}, and the neural-head analyses use commit \NNPackageCommit{}. Reproducibility materials include package code, configurations, aggregate outputs, and public or synthetic examples. The repository includes an end-to-end public demonstration using the TensorFlow Flowers dataset; Appendix~\ref{app:flowers-public-demo} describes this example in detail. Flowers is used to demonstrate the software workflow and is not one of the six applications analyzed in the paper. Restricted datasets and row-level derivatives are not distributed and must be obtained independently under their applicable access terms.

Code and replication materials are available at
\ificlr
   \url{https://anonymous.4open.science/r/lookagain-ml-iclr2027-3504/}.
\else
    \url{https://github.com/vamusesay/lookagain-ml}.
\fi


\section*{Ethics Statement}
The study uses existing medical, demographic, property, agricultural, and auction-image data. Predictions are evaluated as methodological examples and are not intended for diagnosis, valuation, lending, or personnel decisions. Facial and medical applications can encode demographic or institutional biases. We therefore report aggregate predictive performance, avoid causal claims, preserve access restrictions, and do not recommend deployment without domain-specific fairness, privacy, and external-validity review.

\section*{AI Use Statement}
Generative AI tools assisted with code execution, result cross-checking, literature discovery, and drafting. The authors remain responsible for the analysis, citations, claims, and final text, and for the interpretation of the comparisons with saved machine-readable outputs.

\bibliography{references}
\bibliographystyle{iclr2027_conference}

\clearpage
\appendix
\input{appendix/appendix}

\end{document}

%% file: generated/scalars.tex
\newcommand{\HouseResNet}{0.396}
\newcommand{\HouseSelected}{0.629}

\newcommand{\HousePreferred}{0.658}

\newcommand{\HouseGain}{0.029}
\newcommand{\HouseGainLower}{0.017}
\newcommand{\HouseGainUpper}{0.042}

\newcommand{\HorseResNet}{0.029}
\newcommand{\HorseSelected}{0.105}

\newcommand{\HorsePreferred}{0.111}

\newcommand{\HorseGain}{0.006}
\newcommand{\HorseGainLower}{-0.017}
\newcommand{\HorseGainUpper}{0.031}
\newcommand{\HorseRepeatedSelected}{0.102}
\newcommand{\HorseRepeatedPreferred}{-4.009}

\newcommand{\BreakHisSelected}{0.878}

\newcommand{\BreakHisPreferred}{0.931}

\newcommand{\PneumoniaSelected}{0.965}

\newcommand{\PneumoniaPreferred}{0.979}

\newcommand{\AgeSelected}{4.786}

\newcommand{\AgePreferred}{4.867}

\newcommand{\AgeRepeatedSelected}{4.762}
\newcommand{\AgeRepeatedPreferred}{4.638}

\newcommand{\AgeRepeatedSD}{0.094}

\newcommand{\AgeSelectedRMSE}{6.800}
\newcommand{\AgeSelectedCorrelation}{0.961}
\newcommand{\AgeSelectedWithinFive}{65.1\%}
\newcommand{\AgePreferredRMSE}{9.051}
\newcommand{\AgePreferredCorrelation}{0.932}
\newcommand{\AgePreferredWithinFive}{65.1\%}

\newcommand{\LogMEMedianValidationRho}{0.903}
\newcommand{\LogMEWinnerAgreement}{4/6}
\newcommand{\AgeDuplicateFlagged}{8}
\newcommand{\AgeDuplicateTotal}{1465}
\newcommand{\AgeDuplicateMAE}{5.915}
\newcommand{\RiceCloseShare}{37.0\%}
\newcommand{\RiceOneNN}{1.000}
\newcommand{\MatchedHouseGap}{0.178}
\newcommand{\MatchedHorseGap}{0.078}
\newcommand{\MatchedAgeGap}{3.524}
\newcommand{\LightDino}{0.642}
\newcommand{\LightResNet}{0.426}
\newcommand{\LightGap}{0.216}
\newcommand{\MedianNNRankCorrelation}{0.962}
\newcommand{\PackageCommit}{\nolinkurl{877908c916573109b2106cd5fec82ec83a6153de}}

%% file: generated/nn_scalars.tex
\newcommand{\HouseNNGap}{0.225}
\newcommand{\HorseNNGap}{0.071}
\newcommand{\NNPackageCommit}{\nolinkurl{d3d96e09bb49f33dcf3e5cb0a875ac5d661ae0ad}}

%% file: generated/flowers_scalars.tex
\newcommand{\FlowersSelectedValLoss}{0.0397}
\newcommand{\FlowersSelectedAcc}{0.9836}
\newcommand{\FlowersSelectedBalanced}{0.9838}
\newcommand{\FlowersSelectedMacro}{0.9828}
\newcommand{\FlowersSelectedLogLoss}{0.0597}
\newcommand{\FlowersStackValLoss}{0.0448}
\newcommand{\FlowersStackAcc}{0.9836}
\newcommand{\FlowersStackLogLoss}{0.0777}
\newcommand{\FlowersStackAccuracyDifference}{0.0000}
\newcommand{\FlowersStackLogLossDifference}{0.0180}
\newcommand{\FlowersRepeatedSelectedCount}{12}
\newcommand{\FlowersRepeatedStackPreferredCount}{5}
\newcommand{\FlowersRepeatedSelectedMean}{0.9852}
\newcommand{\FlowersRepeatedSelectedSD}{0.0042}
\newcommand{\FlowersRepeatedStackMean}{0.9861}
\newcommand{\FlowersRepeatedStackSD}{0.0037}

%% file: tables/generated_headline.tex
\begin{tabular}{lcccc}
\toprule
Dataset & ResNet & Selected single & Preferred procedure & Gain (95\% CI)\\
\midrule
Houses ($R^2$) & 0.396 & 0.629 (SigLIP 2-B/16) & 0.658 (linear stack) & 0.029 $[0.017,0.042]$\\
Horses ($R^2$) & 0.029 & 0.105 (DINOv2-B/14) & 0.111 (feature union) & 0.006 $[-0.017,0.031]$\\
BreaKHis (acc.) & 0.895 & 0.878 (VGG16) & 0.931 (equal average) & 0.053 $[0.016,0.100]$\\
Pneumonia (acc.) & 0.942 & 0.965 (ConvNeXt-B) & 0.979 (feature union) & 0.015 $[0.002,0.028]$\\
Facial age (MAE) & 9.912 & 4.786 (SigLIP 2-B/16) & 4.867 (linear stack) & -0.081 $[-0.447,0.139]$\\
Rice disease (acc.) & 0.997 & 1.000 (DINOv2-B/14) & 1.000 (feature union) & 0.000 $[0.000,0.000]$\\
\bottomrule
\end{tabular}

%% file: tables/generated_repeated.tex
\begin{tabular}{lccccc}
\toprule
Dataset & Most selected & Single mean & Preferred mean & Mean gain & Preferred SD\\
\midrule
Houses & SigLIP 2-B/16 (11/12) & 0.653 & 0.682 & 0.030 & 0.016\\
Horses & DINOv2-B/14 (12/12) & 0.102 & -4.009 & -4.111 & 14.204\\
BreaKHis & VGG16 (4/12) & 0.823 & 0.882 & 0.060 & 0.042\\
Pneumonia & ConvNeXt-B (6/12) & 0.952 & 0.969 & 0.018 & 0.003\\
Facial age & SigLIP 2-B/16 (12/12) & 4.762 & 4.638 & 0.124 & 0.094\\
Rice disease & DINOv2-B/14 (6/12) & 0.999622 & 0.999348 & -0.000274 & 0.001\\
\bottomrule
\end{tabular}

%% file: appendix/appendix.tex
\section{Dataset Construction, Splits, and Quality Control}
\label{app:dataqc}

Every dataset has a permanent manifest joining an observation identifier, image path, image hash, group identifier, outcome, and split. The cache-validation implementation checks manifest digests, identifiers, and row counts when arrays are available; this does not assert present availability of every original array. Table~\ref{tab:splitqc} summarizes the final locked samples and grouping rules. Independent verification found no group or exact-hash crossing in the fixed split or any of the twelve repeated splits.

\begin{table}[h]
\centering
\small
\caption{Final sample and leakage audit. $N$ is the common-complete analysis sample.}
\label{tab:splitqc}
\begin{tabular}{lrrrl}
\toprule
Dataset & $N$ & Test & Groups & Grouping unit\\
\midrule
Houses & 7,273 & 1,090 & 7,262 & saved / exact hash\\
Horses & 9,729 & 1,459 & 9,461 & reconstructed horse identity\\
BreaKHis & 7,909 & 1,129 & 82 & patient\\
Pneumonia & 5,856 & 878 & 3,117 & patient / exact hash\\
Facial age & 9,764 & 1,465 & 9,658 & exact hash\\
Rice disease & 5,932 & 759 & 4,794 & exact hash\\
\bottomrule
\end{tabular}
\end{table}

The horse grouping is reconstructed conservatively by linking all observations connected through shared identifying information. Exact normalized horse name contributes 19 repeated keys covering 38 rows; complete sire--dam--sex--foaling identity contributes 261 repeated keys covering 522 rows; exact photograph hash contributes 64 repeated keys covering 128 rows. The result has 266 multi-row groups and maximum group size three. Fuzzy names and test-informed links are excluded. Both fixed and repeated assignments have zero crossing identities.

Read-only saved-prediction checks recomputed 5,180 metric comparisons across the six tasks, Flowers, and neural-head mean predictions; 58 checks passed. Additional comparisons covered displayed result rows, cost rows, and scalar definitions. Package tests, manuscript compilation, notebook static validation, and privacy checks have separate records. These checks do not certify that every historical manifest, representation array, fitted model, or generation step was reloaded. 


\subsection{Outcome and task notes}

The Toronto house-price prediction task uses the saved log-price analysis outcome. Horse performance uses \texttt{speed\_z\_within\_sale}, a population z-score of workout speed within sale $\times$ workout-distance groups. BreaKHis classifies benign versus malignant histology images, pneumonia classifies chest radiographs, and rice uses four leaf-disease classes. Facial age uses numerical chronological-age labels as a continuous target. Its primary selection and headline metric is MAE in years; RMSE, correlation, and percentage within five years are secondary. No age-class accuracy or multinomial age stack enters the analysis.

\section{Nested Validation, Selection, and Refit}
\label{app:nested}

Every outer split contains group-disjoint training, validation, and untouched test partitions. Base-head penalties are chosen from group-safe out-of-fold loss inside outer training. The same folds produce one out-of-fold prediction for every outer-training observation and encoder. Base heads are then refitted on all outer training and applied to outer validation.

The combination candidates use the following fitting rules. Equal averaging has no fitted parameters. Linear or multinomial stack weights use only outer-training out-of-fold predictions and outcomes. The greedy subset sequence and stopping point use those same training-only predictions and outcomes. Feature-union scaling, block principal components, and head tuning use a deterministic group-safe inner partition of outer training, after which the candidate is refitted on all outer training. Frozen candidates are compared on outer-validation loss together with the selected single encoder.

In each fixed and repeated split, candidate procedures are constructed and tuned using only the training partition. During the selection stage, validation outcomes are used solely to compare the candidates and choose the preferred procedure. Once selected, all selection and tuning decisions remain fixed. Base heads are then refitted on the combined training and validation data using the selected hyperparameters. If stacking is selected, the stack is refitted using new group-safe out-of-fold predictions generated within the combined training and validation sample. The selected subset remains unchanged, equal averaging remains parameter-free, and feature union is refitted using its previously selected tuning configuration. Test outcomes remain withheld throughout training, tuning, selection, and refitting and are used only for final evaluation.

Near-null directions use the fixed float64 training-only floor $\sqrt{\epsilon}\max(1,\max_j|x_j|)$. A direction below the floor is left unscaled. This prevents division by a numerically null training scale without clipping a large validation or test value.

\section{Complete Fixed-Split Results}
\label{app:fullresults}

\begin{table}[h]
\centering
\small
\caption{Regression single-encoder results. Houses and horses report $R^2$; facial age reports MAE in years, where lower is better.}
\label{tab:regfull}
\input{tables/generated_single_regression}
\end{table}

\begin{table}[h]
\centering
\small
\caption{Classification single-encoder accuracy. Selection uses validation log loss, not the printed test accuracy.}
\label{tab:classfull}
\input{tables/generated_single_classification}
\end{table}

\begin{table}[h]
\centering
\scriptsize
\caption{Fixed-split combination candidates. Each candidate is fitted without outer-validation outcomes. Age is MAE; other columns are $R^2$ or accuracy.}
\label{tab:combofull}
\resizebox{\textwidth}{!}{\input{tables/generated_combinations}}
\end{table}

The selected continuous-age single, SigLIP~2, obtains \AgeSelected{} MAE, \AgeSelectedRMSE{} RMSE, \AgeSelectedCorrelation{} correlation, and \AgeSelectedWithinFive{} within five years. The validation-preferred stack obtains \AgePreferred{} MAE, \AgePreferredRMSE{} RMSE, \AgePreferredCorrelation{} correlation, and \AgePreferredWithinFive{} within five years on the fixed test. Across repeated splits, SigLIP~2 is selected in all twelve; mean single MAE is \AgeRepeatedSelected{} and mean preferred-procedure MAE is \AgeRepeatedPreferred{} (split SD \AgeRepeatedSD{}).

\section{Combination, Diversity, and Sequential Selection}
\label{app:ensembles}

The candidate menu is fixed in the implemented workflow: equal prediction averaging, an all-ten linear or multinomial stack, a greedy subset average, and regularized feature union. Table~\ref{tab:combofull} reports every candidate rather than only the validation winner. Prediction stacking, subset averaging, and feature concatenation are distinct procedures and retain distinct labels throughout.

\begin{figure}[h]
\centering
\includegraphics[width=0.92\textwidth]{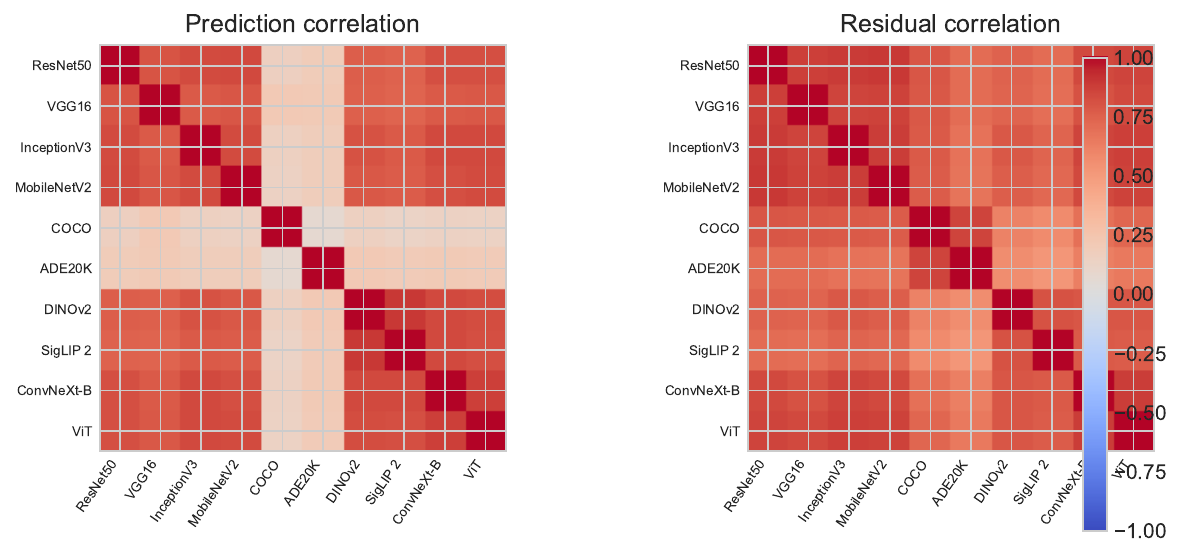}\\[-1mm]
\includegraphics[width=0.92\textwidth]{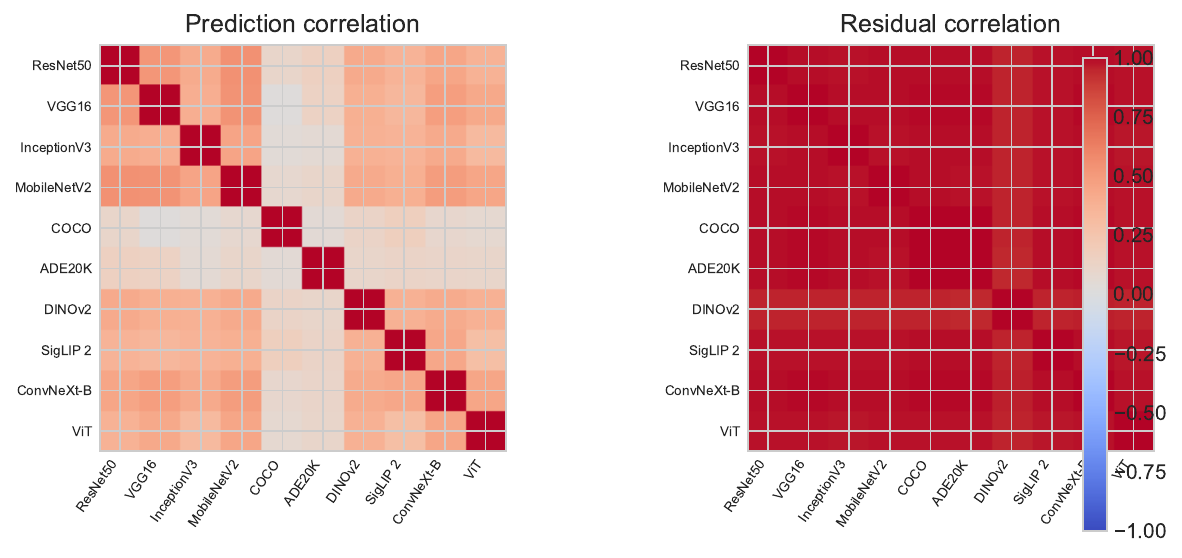}
\caption{\textbf{Prediction and residual correlation diagnostics for houses (top) and horses (bottom).} Each matrix uses the corrected fitted single-encoder predictions on the same locked observations.}
\label{fig:appcorrelations}
\end{figure}

Figure~\ref{fig:appcorrelations} reports corrected fixed-test prediction and residual correlation matrices for houses and horses. Houses retain more residual complementarity among strong encoders. Horse residuals remain highly correlated for many encoder pairs even though prediction correlations vary more. These are descriptive diagnostics, not a universal selection formula. A weak encoder can disagree because it is inaccurate rather than because it contributes useful signal.

\subsection{Horse feature-union instability}


The feature-union failure persists under the group-safe horse design. Union is outer-validation-preferred in three repeated splits. Test $R^2$ is 0.129 in one, $-0.099$ in another, and $-49.112$ in the severe split. 
One possible explanation is an ADE20K feature with little variation in training but extreme values in the test set. Our training-based numerical safeguard does not constrain such test values, and we do not apply test-informed clipping. The severe result demonstrates feature-union instability; it does not establish that prediction stacking necessarily fails.

\section{Dimension Control, Capacity, and Transferability}

The primary common-head design already applies train-only PCA with a 128-dimensional cap. Neural representations above the cap are reduced to 128 directions; COCO's 21 class shares retain their native dimension. Table~\ref{tab:regfull} reports results under this dimension-controlled specification.

The matched DINOv2-B/14 minus supervised ViT-B/16 comparison averages 0.178 $R^2$ for houses and 0.078 for horses over twelve splits. DINOv2 reduces age MAE by 3.524 years. Mean accuracy differences are $-0.017$ for BreaKHis, 0.004 for pneumonia, and 0.001 for rice. Similar parameter scale and output dimension therefore do not account for the whole gap, but the comparison does not isolate a causal training-objective effect.

\begin{table}[h]
\centering
\small
\caption{LogME rank association with validation ranking. Winner agreement is a screening diagnostic and never replaces validation.}
\label{tab:logme}
\input{tables/generated_logme}
\end{table}

LogME winner agreement is \LogMEWinnerAgreement{}. Its median Spearman association with validation ranking is \LogMEMedianValidationRho{}. The low BreaKHis association and rice winner disagreement rule out LogME as a universal selection replacement despite strong associations in most tasks.

\section{Limited Adaptation and Data-Geometry Diagnostics}
\label{app:adaptation}

The house adaptation experiment uses seeds 101, 202, and 303; at most six epochs; AdamW; and encoder and head learning rates $10^{-5}$ and $10^{-4}$. The ResNet50 final residual stage and scalar head are trainable. The DINOv2 final transformer block, final layer normalization, and scalar head are trainable. Earlier encoder parameters are frozen. ResNet50 batch-normalization running statistics remain update-enabled throughout the model because it is in training mode. This separate fixed-split supplemental experiment fits on training observations and uses the fixed validation partition for MSE early stopping, then evaluates both models fixed in the implemented workflow on the untouched test partition without a further same-validation procedure comparison or train-plus-validation refit. Seed-averaged predictions yield $R^2=\LightResNet{}$ for ResNet50 and $\LightDino{}$ for DINOv2. The paired grouped-bootstrap difference is \LightGap{} with 95\% CI $[0.142,0.274]$. SigLIP~2 was not part of the adaptation protocol fixed in the implemented workflow and is not added post hoc.

For continuous age, the DINOv2 nearest-training cosine audit flags 8/1,465 test rows at 0.99 similarity. Removing them changes MAE from 5.908 to 5.915. At 0.995 similarity, 2 rows are flagged and MAE is 5.907; none exceed 0.999. Exact hashes remain the only available identity safeguard, so this is a sensitivity analysis rather than proof of person-level separation.

For rice, 281/759 test observations (37.0\%) have nearest-training cosine similarity at least 0.99, and all share the nearest neighbor's class. A 1-NN classifier obtains 1.000 accuracy, and DINOv2 remains perfect after excluding close-neighbor rows. In a separate locked-split diagnostic, DINOv2 representations were evaluated with linear logistic heads after fitting scaling and PCA on each sampled training set and selecting the penalty by fixed-validation log loss, without a train-plus-validation refit. Across ten training-subsample seeds, mean test accuracy was about 0.936 with 50 images per class, 0.969 with 100, 0.992 with 250, 0.998 with 500, and approximately 0.999 using the full training sample. These results suggest that the rice task is relatively easy because similar images appear throughout the dataset.


\section{Neural-Network Head Ablation}
\label{app:nnfull}

All neural-head experiments reuse the final manifests and cached representations, with the same 128-component cap as the linear heads. Scaling and PCA are fitted within each training partition. A modest architecture grid is selected once per dataset using ResNet50 and DINOv2 as anchors on a group-safe holdout within outer training, then applied to every encoder. Early stopping uses that inner holdout, with MAE for age, squared error for houses and horses, and log loss for classification. Architecture and per-seed epoch budgets are frozen before outer-validation assessment and retained for the train-plus-validation refit. Seeds are fixed at 101, 202, 303, 404, and 505. Headline metrics use predictions averaged across the five seeds; seed SD describes the separate seed-specific metrics, and no best test seed is reported.

\begin{table}[h]
\centering
\small
\caption{Linear and neural heads for the two motivating applications. Values are locked-test $R^2$; neural metrics use predictions averaged across five seeds fixed in the implemented workflow.}
\label{tab:nnmotivation}
\input{tables/generated_nn_motivation}
\end{table}

\begin{table}[h]
\centering
\small
\caption{Continuous-age neural-head MAE and across-seed SD, both in years.}
\label{tab:agenn}
\input{tables/generated_age_nn}
\end{table}

\begin{figure}[h]
\centering
\begin{minipage}{0.58\textwidth}
\includegraphics[width=\linewidth]{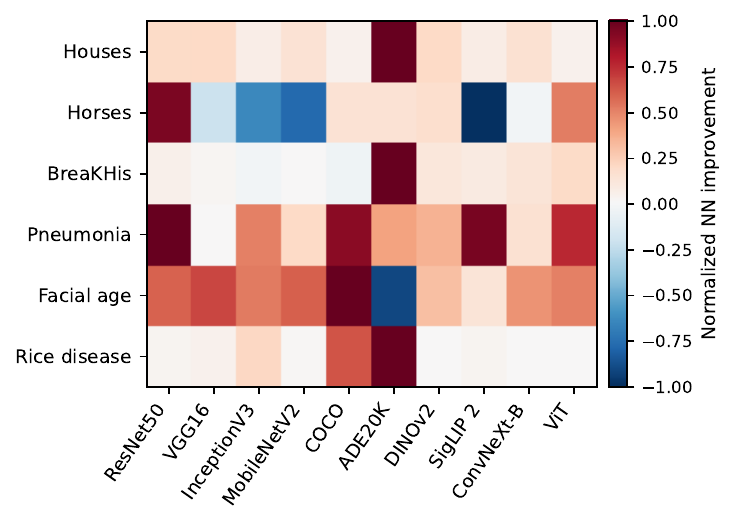}
\end{minipage}\hfill
\begin{minipage}{0.39\textwidth}
\includegraphics[width=\linewidth]{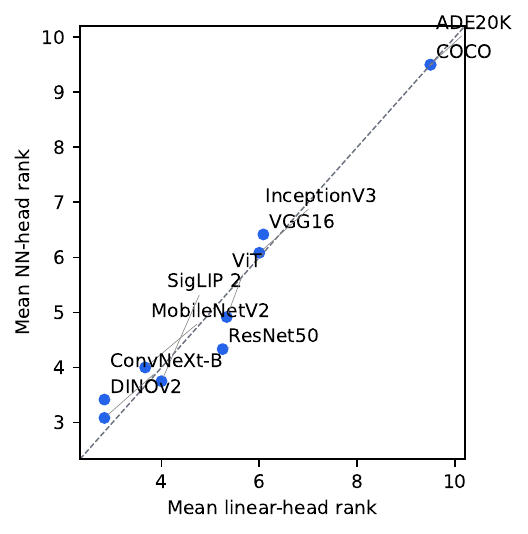}
\end{minipage}
\caption{\textbf{Neural-head robustness.} Left: within-task normalized neural-head improvement for every dataset--encoder cell; positive values favor the neural head. Right: mean linear-head rank against mean neural-head rank across the six tasks. Ties receive average ranks and lower is better.}
\label{fig:appnn}
\end{figure}

Across tasks, the median within-task Spearman correlation between linear and neural encoder rankings is \MedianNNRankCorrelation{}. Nonlinearity helps some representations, but it does not systematically eliminate the ranking differences that motivate encoder comparison.

\section{Computational Cost}
\label{app:cost}

\begin{table}[h]
\centering
\small
\caption{Recorded encoder characteristics and descriptive median cache-creation throughput across seven task records. These heterogeneous timings are not a standardized comparison on common hardware and a common extraction protocol.}
\label{tab:cost}
\input{tables/generated_cost}
\par\smallskip
\begin{minipage}{\textwidth}\footnotesize
\input{generated/cost_note}
\end{minipage}
\end{table}

Initial extraction, cached downstream refitting, and deployment inference are separate costs. Cached representations can be reused for downstream refitting; the available measurements do not establish a universal dominance relation between these costs. Deployment combinations require every constituent encoder, whereas a selected single requires one encoder pass. The recorded extraction measurements depend on hardware, batch size, sample, and I/O. They are not measurements of deployment latency.

\section{Bootstrap and Selection Details}
\label{app:bootstrap}

For fitted procedures $a,b$ and metric $T$, each bootstrap draw samples test groups with replacement, retains all rows in each selected group, and computes
\[
  \Delta^{*(r)}=T\!\left(y^{*(r)},\widehat y_a^{*(r)}\right)
                    -T\!\left(y^{*(r)},\widehat y_b^{*(r)}\right).
\]
The reported interval is the 2.5th and 97.5th percentile of 2,000 paired draws. The same sampled rows enter both procedures. The metrics and their difference are recomputed from fixed fitted predictions; models are not retrained in a draw. The interval therefore conditions on fitting, selection, and the locked split.

Classification procedures minimize validation log loss even when accuracy is printed for interpretability. BreaKHis can therefore select VGG16 despite its lower test accuracy. Test-best columns and regret are descriptive diagnostics and never determine a method.



\input{appendix/flowers_public_demo/public_flowers_demo}

%% file: tables/generated_single_regression.tex
\begin{tabular}{lrrr}
\toprule
Encoder & Houses $R^2$ & Horses $R^2$ & Age MAE\\
\midrule
ResNet50 & 0.3964 & 0.0293 & 9.912\\
VGG16 & 0.3868 & 0.0291 & 11.201\\
InceptionV3 & 0.4200 & 0.0350 & 9.797\\
MobileNetV2 & 0.4195 & 0.0405 & 10.079\\
COCO semantic & 0.0084 & 0.0016 & 20.507\\
ADE20K semantic & -0.1254 & 0.0033 & 28.992\\
DINOv2-B/14 & 0.6207 & 0.1046 & 5.908\\
SigLIP 2-B/16 & 0.6287 & 0.0275 & 4.786\\
ConvNeXt-B & 0.4849 & 0.0279 & 8.934\\
ViT-B/16 & 0.4482 & 0.0195 & 9.298\\
\bottomrule
\end{tabular}

%% file: tables/generated_single_classification.tex
\begin{tabular}{lrrr}
\toprule
Encoder & BreaKHis & Pneumonia & Rice\\
\midrule
ResNet50 & 0.8955 & 0.9419 & 0.9974\\
VGG16 & 0.8778 & 0.9510 & 0.9960\\
InceptionV3 & 0.8379 & 0.9385 & 0.9802\\
MobileNetV2 & 0.9123 & 0.9601 & 0.9987\\
COCO semantic & 0.7281 & 0.7437 & 0.5837\\
ADE20K semantic & 0.5093 & 0.8075 & 0.6469\\
DINOv2-B/14 & 0.8547 & 0.9544 & 1.0000\\
SigLIP 2-B/16 & 0.8778 & 0.9442 & 0.9974\\
ConvNeXt-B & 0.9097 & 0.9647 & 1.0000\\
ViT-B/16 & 0.8645 & 0.9362 & 1.0000\\
\bottomrule
\end{tabular}

%% file: tables/generated_combinations.tex
\begin{tabular}{lrrrrrr}
\toprule
Method & House & Horse & BreaKHis & Pneum. & Age MAE & Rice\\
\midrule
Equal Average & 0.5230 & 0.0573 & 0.9309 & 0.9681 & 9.6181 & 1.0000\\
Linear Stack & 0.6582 & 0.1111 & 0.9300 & 0.9738 & 4.8675 & 1.0000\\
Selected Subset & 0.6615 & 0.1046 & 0.9247 & 0.9772 & 4.7860 & 1.0000\\
Feature Union & 0.5923 & 0.1109 & 0.9123 & 0.9795 & 7.6161 & 1.0000\\
\bottomrule
\end{tabular}

%% file: tables/generated_logme.tex
\begin{tabular}{lrrr}
\toprule
Dataset & Spearman $\rho$ & LogME top & Agrees with validation\\
\midrule
Houses & 0.952 & SigLIP 2-B/16 & yes\\
Horses & 0.855 & DINOv2-B/14 & yes\\
BreaKHis & 0.248 & SigLIP 2-B/16 & no\\
Pneumonia & 0.806 & ConvNeXt-B & yes\\
Facial age & 1.000 & SigLIP 2-B/16 & yes\\
Rice disease & 0.952 & SigLIP 2-B/16 & no\\
\bottomrule
\end{tabular}

%% file: tables/generated_nn_motivation.tex
\begin{tabular}{lrrrr}
\toprule
& \multicolumn{2}{c}{Houses} & \multicolumn{2}{c}{Horses}\\
Encoder & Linear & NN & Linear & NN\\
\midrule
ResNet50 & 0.3964 & 0.4225 & 0.0293 & 0.0349\\
VGG16 & 0.3868 & 0.4143 & 0.0291 & 0.0278\\
InceptionV3 & 0.4200 & 0.4288 & 0.0350 & 0.0312\\
MobileNetV2 & 0.4195 & 0.4395 & 0.0405 & 0.0359\\
COCO semantic & 0.0084 & 0.0139 & 0.0016 & 0.0025\\
ADE20K semantic & -0.1254 & 0.0107 & 0.0033 & 0.0042\\
DINOv2-B/14 & 0.6207 & 0.6479 & 0.1046 & 0.1056\\
SigLIP 2-B/16 & 0.6287 & 0.6385 & 0.0275 & 0.0216\\
ConvNeXt-B & 0.4849 & 0.5054 & 0.0279 & 0.0277\\
ViT-B/16 & 0.4482 & 0.4540 & 0.0195 & 0.0225\\
\bottomrule
\end{tabular}

%% file: tables/generated_age_nn.tex
\begin{tabular}{lrr}
\toprule
Encoder & Test MAE & Seed SD\\
\midrule
ResNet50 & 8.071 & 0.074\\
VGG16 & 9.119 & 0.188\\
InceptionV3 & 8.163 & 0.116\\
MobileNetV2 & 8.197 & 0.040\\
COCO semantic & 17.371 & 0.094\\
ADE20K semantic & 31.845 & 3.111\\
DINOv2-B/14 & 4.962 & 0.047\\
SigLIP 2-B/16 & 4.374 & 0.049\\
ConvNeXt-B & 7.524 & 0.069\\
ViT-B/16 & 7.718 & 0.125\\
\bottomrule
\end{tabular}

%% file: tables/generated_cost.tex
\begin{tabular}{lrrr}
\toprule
Encoder & Parameters (M) & Dimension & Median images/s\\
\midrule
ResNet50 & 23.5 & 2,048 & 166.2\\
VGG16 & 138.4 & 512 & 155.9\\
InceptionV3 & 21.8 & 2,048 & 150.0\\
MobileNetV2 & 2.2 & 1,280 & 171.8\\
COCO semantic & 11.0 & 21 & 180.5\\
ADE20K semantic & 3.8 & 150 & 94.1\\
DINOv2 & 86.6 & 768 & 137.5\\
SigLIP 2 & 92.9 & 768 & 149.9\\
ConvNeXt & 87.6 & 1,024 & 144.5\\
ViT & 85.8 & 768 & 149.7\\
\bottomrule
\end{tabular}

%% file: generated/cost_note.tex
The seven records cover Toronto houses, OBS horses, BreaKHis, pneumonia, facial age, and rice disease;
TensorFlow Flowers is reported separately. Throughput is the median of recorded per-encoder creation images/s, including historical external embeddings and package-created caches, rather than the zero extraction time of a cache-hit analysis. The horse ResNet50 record measures 151 newly extracted images, while other horse encoders measure 9,885 source images; other tasks also differ in sample size and extraction settings. Exact historical hardware equivalence is not established. Parameter counts use the first available Toronto record and its recorded counting convention; they are not medians. These values do not measure deployment latency.

%% file: appendix/flowers_public_demo/public_flowers_demo.tex
\section{Reproducible Public Demonstration with TensorFlow Flowers}
\label{app:flowers-public-demo}
\begingroup
\raggedbottom

This appendix provides a public reproducible workflow used by \pkg{}. It is not a seventh headline application and does not replace the restricted-data analyses. The demonstration asks a narrower question: when observations, labels, downstream head family, and split are held fixed, how much does predictive performance change across defensible frozen representations, and does a combination fixed in the implemented workflow improve on the validation-selected single representation?

\subsection{Data preparation and locked design}

The publicly available TensorFlow Flowers dataset is distributed as a compressed \texttt{.tgz} archive containing 3,670 JPEG images organized into five class directories: daisy, dandelion, roses, sunflowers, and tulips \citep{tfflowers}. The workflow obtains the archive from TensorFlow's official source, verifies its size and SHA-256 digest, safely extracts the images, decodes and content-hashes every file, and builds a permanent observation manifest. Folder names provide class labels, while exact image hashes define duplicate groups that cannot cross partitions. Two cross-class duplicate rows are excluded, leaving 3,668 observations and 3,666 effective groups. The locked group-safe 70/15/15 split contains 2,568/550/550 observations (Table~\ref{tab:flowers-design}).

\input{appendix/flowers_public_demo/tables/study_design}

Within outer training, three-fold group-safe out-of-fold predictions tune each multinomial head and fit the all-ten linear stack. The resulting single-encoder and stack candidates are frozen before their log losses are compared on outer validation. After selection, locked heads are refitted on training plus validation; a stack refit, when selected, uses new group-safe out-of-fold predictions on training plus validation. Test outcomes enter only the final evaluation.

\subsection{Representation comparison}

We compare ten frozen representations under a common multinomial logistic head: ResNet50, VGG16, InceptionV3, MobileNetV2, COCO DeepLab class shares, ADE20K SegFormer class shares, DINOv2-B/14, SigLIP~2-B/16, ConvNeXt-B, and ViT-B/16. The recorded analyses reused cached representations. SigLIP~2-B/16 has the lowest outer-validation log loss, \FlowersSelectedValLoss{}, and is therefore the locked single-model choice (Figure~\ref{fig:flowers-selection}). Its locked-test accuracy is \FlowersSelectedAcc{}, balanced accuracy is \FlowersSelectedBalanced{}, macro F1 is \FlowersSelectedMacro{}, and log loss is \FlowersSelectedLogLoss{} (Table~\ref{tab:flowers-performance}). DINOv2-B/14 is close, whereas the two semantic class-share representations are substantially weaker on this object-classification task. This spread illustrates representation risk; it does not establish a universal encoder ranking.

\begin{figure}[!htbp]
\centering
\includegraphics[width=0.86\linewidth]{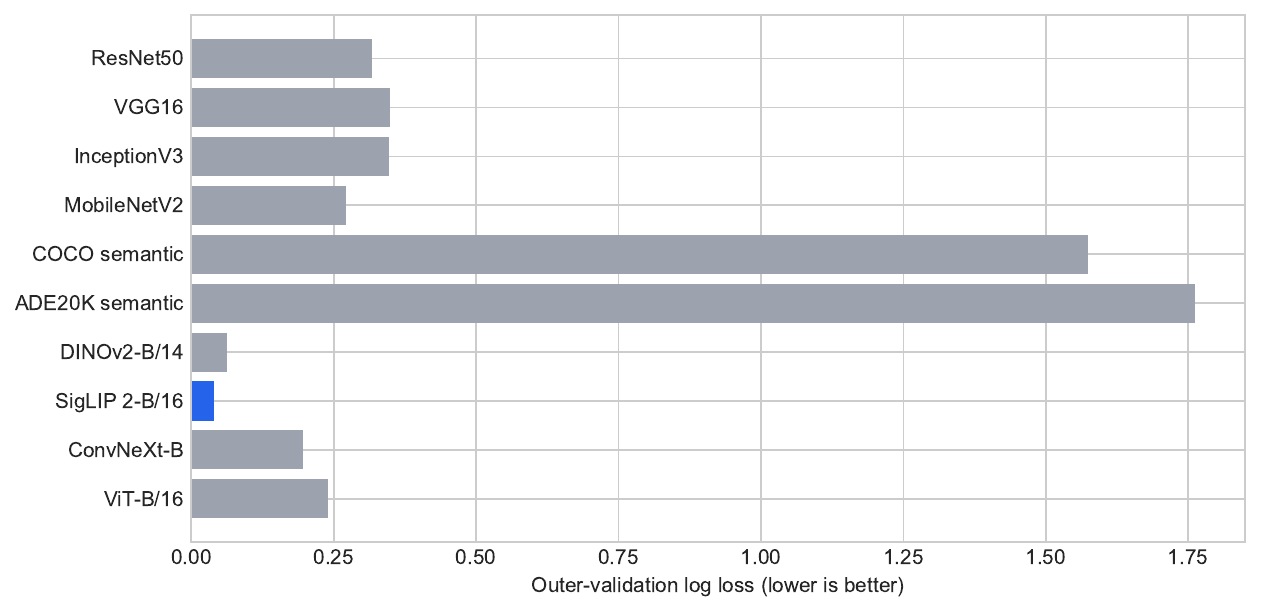}
\caption{Outer-validation log loss for ten frozen representations under training-only tuning. Lower is better; the highlighted representation was selected without using test outcomes.}
\label{fig:flowers-selection}
\end{figure}

\input{appendix/flowers_public_demo/tables/encoder_performance}

\subsection{Combination and paired uncertainty}

The all-ten linear stack is fitted from outer-training out-of-fold predictions and outcomes. Its outer-validation log loss is \FlowersStackValLoss{}, compared with \FlowersSelectedValLoss{} for SigLIP~2, so the preferred primary procedure remains the selected single encoder. The frozen stack nevertheless provides a useful descriptive comparison: its locked-test accuracy is \FlowersStackAcc{} and its log loss is \FlowersStackLogLoss{}. Stack-minus-single point differences are \FlowersStackAccuracyDifference{} for accuracy and \FlowersStackLogLossDifference{} for log loss (Table~\ref{tab:flowers-uncertainty}; Figure~\ref{fig:flowers-uncertainty}).

\input{appendix/flowers_public_demo/tables/paired_uncertainty}

\begin{figure}[!htbp]
\centering
\includegraphics[width=0.82\linewidth]{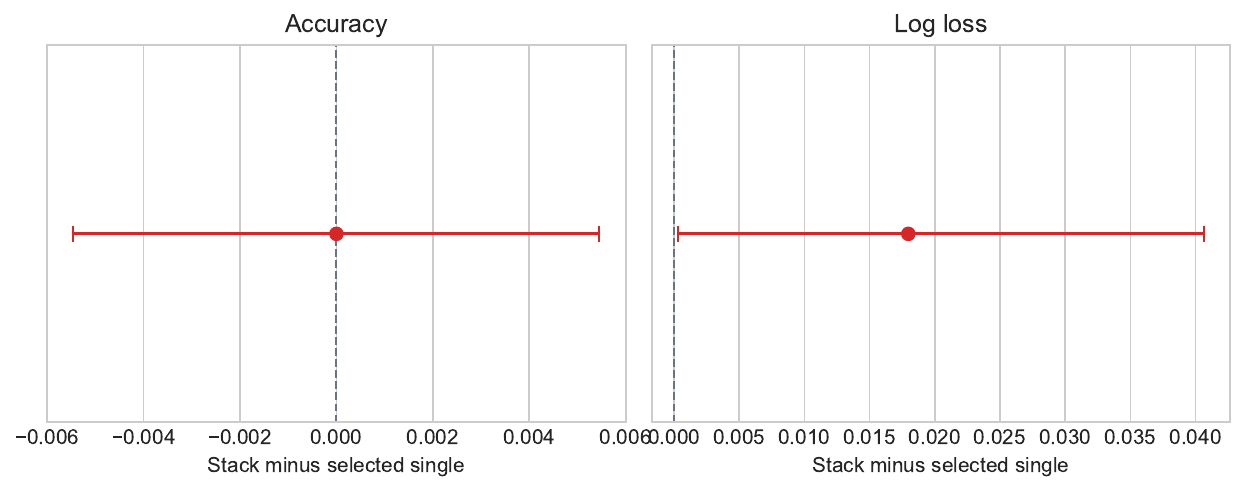}
\caption{All-ten stack minus validation-selected SigLIP~2-B/16 on the locked test. Bars are paired 95\% conditional bootstrap intervals from fixed fitted predictions. Zero denotes no difference; negative log-loss differences favor the stack.}
\label{fig:flowers-uncertainty}
\end{figure}

The percentile intervals use 2,000 paired, group-aware test resamples. Both performance metrics and their difference are recomputed from the fixed fitted predictions; models are not retrained inside a draw. The intervals condition on fitting, selection, and the locked split.

\subsection{Supplementary split stability}

As a supplementary exercise, we repeat the group-safe split twelve times using seeds fixed in the implemented workflow. Representations are not recomputed. Within each split, base heads and the stack are reconstructed from training-only out-of-fold predictions, frozen candidates are compared on outer validation, and test outcomes are evaluated only after selection. SigLIP~2-B/16 is the selected single encoder in \FlowersRepeatedSelectedCount{} of 12 splits and has mean test accuracy \FlowersRepeatedSelectedMean{} (SD \FlowersRepeatedSelectedSD{}). The stack is the preferred procedure in \FlowersRepeatedStackPreferredCount{} splits and has mean test accuracy \FlowersRepeatedStackMean{} (SD \FlowersRepeatedStackSD{}). These empirical ranges measure partition sensitivity and are not confidence intervals (Table~\ref{tab:flowers-stability}; Figure~\ref{fig:flowers-stability}).

\input{appendix/flowers_public_demo/tables/split_stability}

\begin{figure}[!htbp]
\centering
\includegraphics[width=0.86\linewidth]{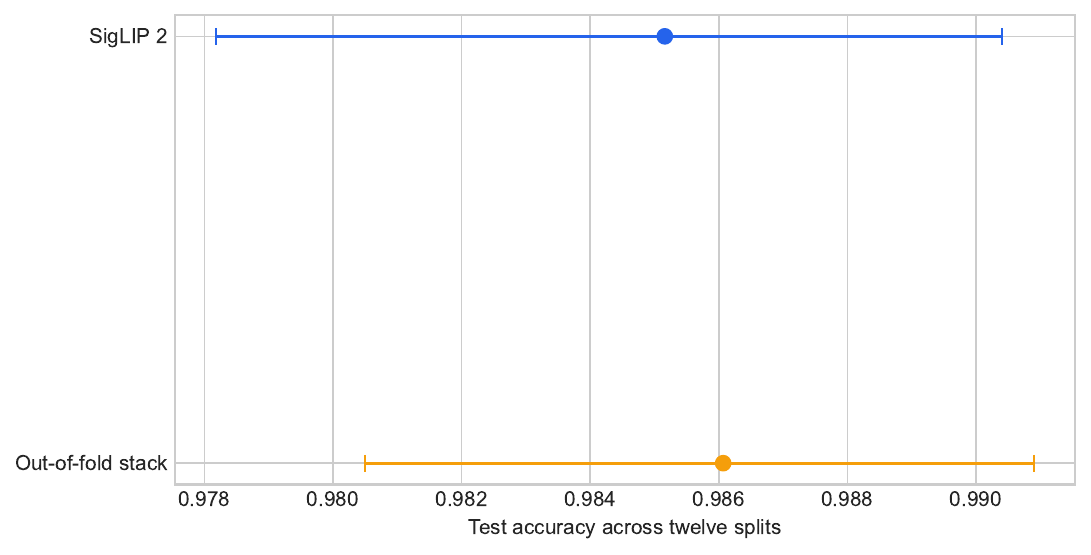}
\caption{Empirical test-accuracy variation across twelve group-safe splits fixed in the implemented workflow for the selected single encoder and all-ten cross-fitted stack. Bars show empirical 2.5th--97.5th percentiles, not confidence intervals.}
\label{fig:flowers-stability}
\end{figure}

\subsection{Computation, reproducibility, and interpretation}

On the replication machine's NVIDIA RTX 3060, verified representation-creation times ranged from 20.5 to 84.6 seconds for the 3,668-image analytical sample, excluding model-download time (Table~\ref{tab:flowers-cost}; Figure~\ref{fig:flowers-cost}). These timings are device- and I/O-specific. The corrected fixed and twelve-split analyses reused all ten verified representation caches and ran through \pkg{} commit \PackageCommit{}.

\input{appendix/flowers_public_demo/tables/computational_cost}

\begin{figure}[!htbp]
\centering
\includegraphics[width=0.86\linewidth]{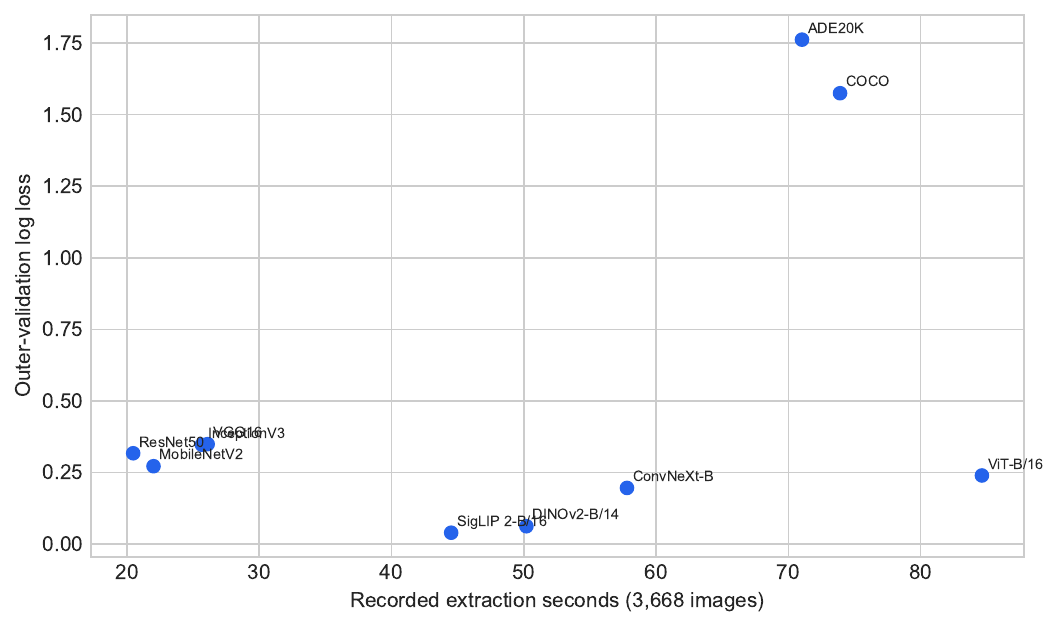}
\caption{Observed representation extraction time and outer-validation log loss on the recorded analysis environment. Timings describe this hardware and sample and should not be generalized mechanically.}
\label{fig:flowers-cost}
\end{figure}

This example is included to demonstrate the software workflow rather than as an additional empirical application.

\endgroup

%% file: appendix/flowers_public_demo/tables/study_design.tex
\begin{table}[!htbp]
\centering
\small
\caption{Public TensorFlow Flowers demonstration: sample construction and locked design.}
\label{tab:flowers-design}
\begin{tabular}{lr}
\toprule
Quantity & Value \\
\midrule
JPEG images in the source archive & 3,670 \\
Excluded cross-class duplicate rows & 2 \\
Analytical observations & 3,668 \\
Duplicate-safe groups & 3,666 \\
Train / validation / locked test & 2,568 / 550 / 550 \\
Supplementary group-safe splits & 12 \\
Paired locked-test bootstrap draws & 2,000 \\
\bottomrule
\end{tabular}
\end{table}

%% file: appendix/flowers_public_demo/tables/encoder_performance.tex
\begin{table}[!htbp]
\centering\small
\caption{Locked-test Flowers performance after training-only tuning. Selection uses outer-validation log loss.}
\label{tab:flowers-performance}
\begin{tabular}{lrrrr}
\toprule
Representation & Accuracy & Balanced acc. & Macro F1 & Log loss\\
\midrule
ResNet50 & 0.8909 & 0.8907 & 0.8897 & 0.3658\\
VGG16 & 0.8855 & 0.8866 & 0.8860 & 0.3421\\
InceptionV3 & 0.8782 & 0.8761 & 0.8768 & 0.3428\\
MobileNetV2 & 0.9091 & 0.9097 & 0.9085 & 0.3197\\
COCO semantic & 0.3127 & 0.2986 & 0.2649 & 1.5285\\
ADE20K semantic & 0.4200 & 0.4001 & 0.3915 & 1.5292\\
DINOv2-B/14 & 0.9818 & 0.9821 & 0.9815 & 0.0658\\
SigLIP 2-B/16 & 0.9836 & 0.9838 & 0.9828 & 0.0597\\
ConvNeXt-B & 0.9382 & 0.9381 & 0.9373 & 0.2236\\
ViT-B/16 & 0.9364 & 0.9372 & 0.9360 & 0.2068\\
\bottomrule
\end{tabular}
\end{table}

%% file: appendix/flowers_public_demo/tables/paired_uncertainty.tex
\begin{table}[!htbp]
\centering\small
\caption{Paired Flowers comparison from fixed fitted test predictions. Differences are stack minus selected single.}
\label{tab:flowers-uncertainty}
\begin{tabular}{lrrr}
\toprule
Metric & SigLIP~2 & Out-of-fold stack & Difference (95\% CI)\\
\midrule
Accuracy & 0.9836 & 0.9836 & 0.0000 $[-0.0055,0.0055]$\\
Log loss & 0.0597 & 0.0777 & 0.0180 $[0.0003,0.0406]$\\
\bottomrule
\end{tabular}
\end{table}

%% file: appendix/flowers_public_demo/tables/split_stability.tex
\begin{table}[!htbp]
\centering\small
\caption{Flowers split stability under nested out-of-fold fitting and outer-validation selection.}
\label{tab:flowers-stability}
\begin{tabular}{lrrrl}
\toprule
Procedure & Mean acc. & SD & Empirical 95\% range & Selection frequency\\
\midrule
SigLIP~2 & 0.9852 & 0.0042 & $[0.9782,0.9904]$ & 12/12 selected\\
All-ten cross-fitted stack & 0.9861 & 0.0037 & $[0.9805,0.9909]$ & 5/12 preferred\\
\bottomrule
\end{tabular}
\end{table}

%% file: appendix/flowers_public_demo/tables/computational_cost.tex
\begin{table}[!htbp]
\centering
\small
\caption{Fresh encoder extraction on the replication machine (RTX 3060, 3,668 images). Timings exclude model download.}
\label{tab:flowers-cost}
\begin{tabular}{lrrr}
\toprule
Encoder & Seconds & Images/s & Peak GPU GiB\\
\midrule
ResNet50 & 20.5 & 179.3 & 0.17\\
VGG16 & 26.1 & 140.6 & 0.81\\
InceptionV3 & 25.7 & 143.0 & 0.20\\
MobileNetV2 & 22.0 & 166.7 & 0.09\\
COCO DeepLab & 73.9 & 49.6 & --\\
ADE20K SegFormer & 71.0 & 51.6 & --\\
DINOv2-B/14 & 50.2 & 73.1 & 0.42\\
SigLIP 2-B/16 & 44.5 & 82.4 & 1.46\\
ConvNeXt-B & 57.8 & 63.5 & 0.46\\
ViT-B/16 & 84.6 & 43.3 & 0.40\\
\bottomrule
\end{tabular}
\end{table}